\documentclass[12pt]{article}

\author{Yu.~M.~Zinoviev
       \thanks{E-mail address: ZINOVIEV@MX.IHEP.SU} \\
        {\it Institute for High Energy Physics} \\
        {\it Protvino, Moscow Region, 142280, Russia}}
\title{First Order Formalism \\
       for Mixed Symmetry Tensor Fields }

\date{}

\begin{document}

\maketitle

\begin{abstract}
In this paper we give explicit first order Lagrangian formulation for
mixed symmetry tensor fields $\Phi_{[\mu\nu],\alpha}$,
$T_{[\mu\nu\alpha],\beta}$ and $R_{[\mu\nu],[\alpha\beta]}$. We show
that such Lagrangians could be written in a very suggestive form
similar to the well known tetrad formalism in gravity. Such
description could simplify the investigations of possible interactions
for these fields. Some examples of interactions are given.
\end{abstract}

\newpage
\setcounter{page}{1}

\section*{Introduction}

Last times there is a renewed interest in the mixed symmetry high spin
tensor fields \cite{Cur86}-\cite{Zin02}. The reason is that such
fields naturally appear in a number of physically interesting theories
such as superstrings, supergravities and (supersymmetric) high spin
theories. One of the technical problems one faces working with such
fields is that to get analog of gauge invariant "field strengths" one
has to build expressions with more and more derivatives \cite{BB02},
\cite{MH02}. This problem appears already for the massless spin two
field in gravity, but in this case there exist very elegant solution:
one goes to the first order formalism and obtains the description in
terms of tetrad $e_\mu{}^a$ and Lorentz connection
$\omega_\mu{}^{[ab]}$ which play the roles of the gauge fields and
"normal" gauge invariant field strengths $T_{[\mu\nu]}{}^a$ and
$R_{[\mu\nu]}{}^{[ab]}$ having nice geometrical interpretation as
torsion and curvature.

In this paper we follow the same way for the three examples of
massless mixed symmetry tensor fields $\Phi_{[\mu\nu],\alpha}$,
$T_{[\mu\nu\alpha],\beta}$ and $R_{[\mu\nu],[\alpha\beta]}$. In all
three cases we managed to construct first order formalism which is
very much similar to the tetrad formalism in gravity and in which the
Lagrangians take very simple and suggestive form. We hope that such
formalism could simplify the investigation of possible interactions
among such fields. Moreover, we give a couple of simple but
interesting examples of such interactions.

For completeness, we remind basic properties of the first order
formalism in gravity which will be important for us throughout the
paper. Usual second order formulation for the massless spin two field
uses symmetric second rank tensor field $h_{\{\mu\nu\}}$ with the
Lagrangian:
\begin{equation}
{\cal L}_0 = \frac{1}{2} \partial_\alpha h^{\mu\nu} \partial_\alpha
h_{\mu\nu} - (\partial h)^\mu (\partial h)_\mu + (\partial h)^\mu
\partial_\mu h - \frac{1}{2} \partial^\mu h \partial_\mu h
\end{equation}
where $(\partial h)^\mu = \partial_\nu h^{\nu\mu}$, $h = h_\mu{}^\mu$,
which is invariant under the gauge transformations
$\delta h_{\mu\nu} = \partial_\mu \xi_\nu + \partial_\nu \xi_\mu$.
In this, there is no any gauge invariant expression with one
derivative analogous to field strength in Yang-Mills theories and the
Lagrangian can not be rewritten as expression quadratic in them.
As is well known the simplest gauge invariant object is the tensor:
$$
R_{[\mu\nu],[\alpha\beta]} = \partial_\mu \partial_\alpha h_{\nu\beta}
- \partial_\nu \partial_\alpha h_{\mu\beta} - \partial_\mu
\partial_\beta h_{\nu\alpha} + \partial_\nu \partial_\beta
h_{\mu\alpha}
$$

Now, let us abandon the requirement that the field $h_{\mu\nu}$ is
symmetric. Instead, let us consider the most general second order
Lagrangian and require that it will be invariant under the gauge
transformations $\delta h_{\mu\nu} = \partial_\mu \xi_\nu$. This time,
it is a trivial task to construct gauge invariant object which is
first order in derivatives: $T_{[\mu\nu],\alpha} = \partial_\mu
h_{\nu\alpha} - \partial_\nu h_{\mu\alpha}$  so that the most general
gauge invariant Lagrangian could be written as:
\begin{equation}
{\cal L} = \frac{1}{8} T^{\mu\nu,\alpha} T_{\mu\nu,\alpha} +
\frac{a_1}{2} T^{\mu\nu,\alpha} T_{\mu\alpha,\nu} + \frac{a_2}{2}
T^\mu T_\mu
\end{equation}
where $T_\mu = T_{\mu\nu}{}^\nu$. But in so doing we have introduced
additional degrees of freedom, namely antisymmetric part of
$h_{\mu\nu}$. Let us decompose our field into symmetric and
antisymmetric parts: $h_{\mu\nu} = f_{\{\mu\nu\}} + B_{[\mu\nu]}$. In
terms of $f_{\mu\nu}$ and $B_{\mu\nu}$ our Lagrangian takes the form:
\begin{eqnarray}
{\cal L} &=& \frac{1+2 a_1}{4} \partial^\mu h^{\nu\alpha} \partial_\mu
h_{\nu\alpha} - \frac{1+2 a_1-2 a_2}{4} (\partial f)^\mu
(\partial f)_\mu - a_2 (\partial f)^\mu \partial_\mu f + \frac{a_2}{2}
\partial^\mu f \partial_\mu f + \nonumber \\
 && + \frac{1-2 a_1}{4} \partial^\mu B^{\nu\alpha} \partial_\mu
 B_{\nu\alpha} - \frac{1-6 a_1-2 a_2}{4} (\partial B)^\mu
 (\partial B)_\mu - \nonumber \\
&& - \frac{1+2 a_1+2 a_2}{2} (\partial f)^\mu  (\partial B)_\mu
\end{eqnarray}
We see that in general there exists mixing between $f_{\mu\nu}$ and
$B_{\mu\nu}$ components so to understand which physical degrees of
freedom such Lagrangian describes one has to go through careful
Hamiltonian analysis. But there exists at least one trivial solution
\footnote{Note that in this the Lagrangian written in terms of "
torsion" $T_{\mu\nu,\alpha}$ belongs to one parameter family of so
called teleparallel Lagrangians.}: $2 a_2 = -( 1 + 2 a_1)$ for
which mixing term is absent and we get:
\begin{eqnarray}
{\cal L} &=& \frac{1+2 a_1}{2} \left[ \frac{1}{2} \partial^\mu
h^{\nu\alpha} \partial_\mu h_{\nu\alpha} - (\partial f)^\mu
(\partial f)_\mu + (\partial f)^\mu \partial_\mu f - \frac{1}{2}
\partial^\mu f \partial_\mu f \right] \nonumber \\
 && + \frac{1-2 a_1}{4} \left[ \partial^\mu B^{\nu\alpha} \partial_\mu
 B_{\nu\alpha} - 2 (\partial B)^\mu  (\partial B)_\mu \right]
\end{eqnarray}
One can see that up to normalization we have a sum of two independent
Lagrangians: usual gauge invariant Lagrangian for symmetric tensor as
well as (also gauge invariant) Lagrangian for antisymmetric field.
The well known tetrad formulation of gravity (i.e. massless spin two
filed) corresponds to the choice $a_1 = 1/2$. In this, antisymmetric
field $B_{\mu\nu}$ completely decouples and does not introduce any
additional physical degrees of freedom, while the Lagrangian has the
form:
\begin{equation}
{\cal L} = \frac{1}{8} T^{\mu\nu,\alpha} T_{\mu\nu,\alpha} +
\frac{1}{4} T^{\mu\nu,\alpha} T_{\mu\alpha,\nu} - \frac{1}{2}
T^\mu T_\mu \label{eq1}
\end{equation}
It is easy to check that this Lagrangian is invariant not only under
the gauge transformations $\delta h_{\mu\nu} = \partial_\mu \xi_\nu$,
but under the local shifts $h_{\mu\nu} \rightarrow h_{\mu\nu} +
\eta_{[\mu\nu]}$ as well\footnote{Clearly, this is a manifestation of
$B_{\mu\nu}$ decoupling}. This last invariance suggests the following
procedure for transition from second order formalism to the first
order one. One introduces auxiliary field $\omega_{\mu,[\alpha\beta]}$
which will play the role of gauge field for the local $\eta_{\mu\nu}$
transformations. Then it is easy to construct the first order
Lagrangian
\begin{equation}
{\cal L}_I = - \frac{1}{2} \omega^{\mu,\alpha\beta}
\omega_{\alpha,\mu\beta} + \frac{1}{2} \omega^\mu \omega_\mu -
\frac{1}{2} \omega^{\mu,\alpha\beta} T_{\alpha\beta,\mu} - \omega^\mu
T_\mu \label{eq2}
\end{equation}
which is invariant under the gauge transformations
$\delta \omega_{\mu,\alpha\beta} = \partial_\mu \eta_{\alpha\beta}$,
$\delta h_{\mu\nu} = \eta_{\mu\nu}$ as well as (trivially) under the
$\delta h_{\mu\nu} = \partial_\mu \xi_\nu$. The algebraic equations of
motion for the $\omega$-field can be solved giving us:
\begin{equation}
\omega_{\mu,\alpha\beta} = \frac{1}{2} [ T_{\mu\alpha,\beta} -
T_{\mu\beta,\alpha} - T_{\alpha\beta,\mu} ]
\end{equation}
In this, substituting this expression into the first order Lagrangian
(\ref{eq2}) one obtains exactly the second order Lagrangian
(\ref{eq1}).

Moreover, if we formally divide all indices into the "local" and
"world" ones this Lagrangian could be rewritten in a very simple and
suggestive form:
\begin{equation}
{\cal L}_I = \frac{1}{2} \left\{ \phantom{|}^{\mu\nu}_{ab}
 \right\} \omega_\mu{}^{ac} \omega_\nu{}^{bc} - \frac{1}{2}
\left\{ \phantom{|}^{\mu\nu\alpha}_{abc} \right\}
\omega_\mu{}^{ab} \partial_\nu h_\alpha{}^c \label{eq3}
\end{equation}
Here
$$
\left\{ \phantom{|}^{\mu\nu}_{ab} \right\} = \delta_a{}^\mu
\delta_b{}^\nu - \delta_a{}^\nu \delta_b{}^\mu
$$
and so on. Such form points to the nice geometric interpretation of
these fields we have already mentioned and greatly simplify
understanding of the possible interactions. Indeed, if one suppose
that interaction Lagrangian has the same general form as the free one,
then one easily, by the finite number of iterations can reproduce
complete non-linear gravitational interactions.

As it is evident from (\ref{eq3}) this construction works for the
space-time dimensions $d \ge 3$ only. In the minimal
dimension\footnote{there are no physical degrees of freedom in this
case} $d = 3$ one can introduce dual variables:
\begin{equation}
f_\mu{}^a = \frac{1}{2} \varepsilon^{abc} \omega_{\mu,bc}, \qquad
\eta^a = \frac{1}{2} \varepsilon^{abc} \eta_c
\end{equation}
and rewrite the Lagrangian in the following simple form:
\begin{equation}
{\cal L} = - \frac{1}{2} \left\{ \phantom{|}^{\mu\nu}_{ab} \right\}
f_\mu{}^a f_\nu{}^b + \varepsilon^{\mu\nu\alpha} f_\mu{}^a
\partial_\nu h_\alpha{}^a
\end{equation}
in this, gauge transformations look like:
$$
\delta f_\mu{}^a = \partial_\mu \eta^a \qquad \delta h_{\mu a} = -
\varepsilon_{\mu a b} \eta^b
$$
Then one can introduces rather exotic interaction \cite{BG00} by
adding to the Lagrangian and gauge transformations additional terms:
\begin{equation}
{\cal L}_{int} = \frac{\kappa}{6} \varepsilon^{\mu\nu\alpha}
\varepsilon_{abc} f_\mu{}^a f_\nu{}^b f_\alpha{}^c \qquad \delta_1
h_{\alpha b} = \kappa \varepsilon_{abc} f_\alpha{}^a \eta^c
\end{equation}
where $\kappa$ --- arbitrary constant having dimension $m^{-2}$.
Note that in the first order formalism such a model looks very simple,
but if one tries to solve algebraic equation of motion for the
$f$-field, one obtains essentially non-linear theory with infinite
number of derivatives.

In the following three sections we consider generalization of this
approach for massless mixed symmetry tensor fields
$\Phi_{[\mu\nu],\alpha}$, $T_{[\mu\nu\alpha],\beta}$ and
$R_{[\mu\nu],[\alpha\beta]}$.

\section{$\Phi_{[\mu\nu],\alpha}$ tensor}

Usual second order formulation for this field uses the tensor
$\Phi_{[\mu\nu],\alpha}$ satisfying additional condition
$\Phi_{[\mu\nu,\alpha]} = 0$. In this, the free Lagrangian is
invariant under two gauge transformations:
$$
\delta \Phi_{\mu\nu,\alpha} = \partial_\mu x_{\nu\alpha} -
\partial_\nu x_{\mu\alpha} + 2 \partial_\alpha y_{\mu\nu} -
\partial_\mu y_{\nu\alpha} + \partial_\nu y_{\mu\alpha}
$$
where $x_{\{\mu\nu\}}$ is symmetric while $y_{[\mu\nu]}$
antisymmetric. As in the spin two case to construct a gauge invariant
object one needs at least two derivatives. Let us abandon additional
condition $\Phi_{[\mu\nu,\alpha]} = 0$ and simultaneously join two
gauge transformations into:
$\delta \Phi_{\mu\nu,\alpha} = \partial_\mu z_{\nu\alpha} -
\partial_\nu z_{\mu\alpha}$
where $z_{\mu\nu}$ is arbitrary second rank tensor \cite{BCNS02},
\cite{BCCSS03}. Then it is easy to get gauge invariant field strength
with one derivative:
$$
T_{\mu\nu\alpha,\beta} = \partial_{[\mu} \Phi_{\nu\alpha],\beta}
$$
Let us consider the most general second order free Lagrangian
invariant under the $z_{\mu\nu}$-transformations:
\begin{equation}
{\cal L} = - \frac{1}{6} T^{\mu\nu\alpha,\beta} T_{\mu\nu\alpha,\beta}
+ \frac{a_1}{4} T^{\mu\nu\alpha,\beta} T_{\mu\nu\beta,\alpha} +
\frac{a_2}{4} T^{\mu\nu} T_{\mu\nu}
\end{equation}
In discarding additional condition $\Phi_{[\mu\nu,\alpha]} = 0$ we
introduce additional degrees of freedom into the theory, so to
understand what physical degrees of freedom such a model describes we
decompose our field
$\Phi_{\mu\nu,\alpha} = \hat{\Phi}_{\mu\nu,\alpha} +
C_{\mu\nu\alpha}$, where $\hat{\Phi}_{[\mu\nu,\alpha]} = 0$ and
$C_{[\mu\nu\alpha]}$ --- completely antisymmetric tensor. In terms of
these components the Lagrangian takes the form:
\begin{eqnarray}
{\cal L} &=& - \frac{2-a_1}{4} \partial^\mu
\hat{\Phi}^{\alpha\beta,\nu} \partial_\mu \hat{\Phi}_{\alpha\beta,\nu}
+ \frac{a_2}{4} \partial_\mu \hat{\Phi}^{\alpha\beta,\mu} \partial_\nu
\hat{\Phi}_{\alpha\beta,\nu} + \frac{2-a_1}{2} \partial_\mu
\hat{\Phi}^{\mu\alpha,\beta} \partial^\nu \hat{\Phi}_{\nu\alpha,\beta}
+ \nonumber \\
 && + a_2 \partial_\mu \hat{\Phi}^{\alpha\beta,\mu} \partial_\alpha
\hat{\Phi}_\beta + \frac{a_2}{2} \partial^\alpha \hat{\Phi}^\beta
\partial_\alpha \hat{\Phi}_\beta - \frac{a_2}{2}
(\partial \hat{\Phi})  (\partial \hat{\Phi}) - \nonumber \\
 && - \frac{1+a_1}{2} \partial^\mu C^{\nu\alpha\beta} \partial_\mu
C_{\nu\alpha\beta} + \frac{4+7a_1+a_2}{4}
(\partial C)_{\alpha\beta}{}^2 + \nonumber \\
 && + \frac{a_1+a_2-2}{2} \partial^\mu
\hat{\Phi}^{\alpha\beta,\mu} (\partial C)_{\alpha\beta}
\end{eqnarray}
Once again we see that there is a mixing of two components so to
understand what degrees of freedom such theory describes and how
physical they are one needs to go through careful analysis. But there
is one evident solution, namely, $a_2 = 2 - a_1$. In this, the
Lagrangian becomes a sum of two independent Lagrangians for the fields
$\hat{\Phi}_{\mu\nu,\alpha}$ and $C_{\mu\nu\alpha}$. Such a theory
could be of some interest, especially in $d=4$ where
$\hat{\Phi}_{\mu\nu,\alpha}$ does not describe any physical degrees
of freedom, while $C_{\mu\nu\alpha}$ is equivalent to the vector
field. Following the tetrad formulation of gravity in this paper we
will adopt the most conservative approach "one field --- one object"
and we will choose $a_1 = -1$, $a_2 = 3$. Then it is easy to check
that the resulting Lagrangian
\begin{equation}
{\cal L} = - \frac{1}{6} T^{\mu\nu\alpha,\beta} T_{\mu\nu\alpha,\beta}
- \frac{1}{4} T^{\mu\nu\alpha,\beta} T_{\mu\nu\beta,\alpha} +
\frac{3}{4} T^{\mu\nu} T_{\mu\nu}
\end{equation}
is invariant not only under the $z$-transformations, but under
the local shifts $\delta \Phi_{\mu\nu,\alpha} = \eta_{[\mu\nu\alpha]}$
as well. This, in turn, suggests the following procedure for
transition to first order formalism. One introduces auxiliary field
$\Omega_{\mu,[\nu\alpha\beta]}$ antisymmetric in the last three
indices which will play a role of gauge field for the
$\eta$-transformations. Then it is easy to construct a first order
Lagrangian
\begin{equation}
{\cal L} = \frac{3}{4} \Omega^{\mu,\nu\alpha\beta}
\Omega_{\nu,\mu\alpha\beta} - \frac{3}{4} \Omega^{\alpha\beta}
\Omega_{\alpha\beta} - \frac{1}{2} \Omega^{\mu,\nu\alpha\beta}
T_{\nu\alpha\beta,\mu} + \frac{3}{2} \Omega^{\alpha\beta}
T_{\alpha\beta}
\end{equation}
invariant both under the $\delta \Phi_{\mu\nu,\alpha} = \partial_\mu
z_{\nu\alpha} - \partial_\nu z_{\mu\alpha}$ and
\begin{equation}
\delta \Omega_\mu{}^{\nu\alpha\beta} = \partial_\mu
\eta^{\nu\alpha\beta} \qquad \delta \Phi_{\mu\nu,\alpha} =
\eta_{\mu\nu\alpha}
\end{equation}

Now, if we solve the algebraic equation of motion for the $\Omega$
field, we get:
\begin{equation}
\Omega_{\mu,\nu\alpha\beta} = \frac{2}{3} T_{\nu\alpha\beta,\mu} +
\frac{1}{3} [ T_{\mu\alpha\beta,\nu} + T_{\mu\nu\alpha,\beta} +
T_{\mu\beta\nu,\alpha} ]
\end{equation}
Substituting this expression back into the first order Lagrangian
gives us exactly the second order one. Moreover, if we formally
divide all indices into the "local" and "world" ones, then this
Lagrangian could also be rewritten in a very simple and suggestive
form:
\begin{equation}
{\cal L} = - \frac{1}{4} \left\{
\phantom{|}^{\mu\nu\alpha\beta}_{abcd}
\right\} [ 3 \Omega_\mu{}^{aef} \Omega_\nu{}^{bef} \delta_\alpha{}^c
\delta_\beta{}^d - \frac{1}{3} \Omega_\mu{}^{abc}
T_{\nu\alpha\beta}{}^d ]
\end{equation}

As we see from the last formula the whole construction works for the
space-time dimensions $d \ge 4$ only. In $d = 4$ (where
$\Phi_{\mu\nu,\alpha}$ does not describe any physical degrees of
freedom) one can introduce dual variables:
$$
f_\mu{}^a = \frac{1}{6} \varepsilon^{abcd} \Omega_\mu{}^{bcd}, \qquad
\eta^a = \frac{1}{6} \varepsilon^{abcd} \eta_{bcd}
$$
Then the first order Lagrangian could be rewritten in the following
form:
\begin{equation}
{\cal L} = - \frac{1}{2} \varepsilon^{\mu\nu\alpha\beta} [ 3
\varepsilon_{abcd} f_\mu{}^a f_\nu{}^b \delta_\alpha{}^c
\delta_\beta{}^d - f_\mu{}^a T_{\nu\alpha\beta}{}^a ]
\end{equation}
in this, the gauge transformations look like:
$$
\delta f_\mu{}^a = \partial_\mu \eta^a \qquad \delta \Phi_{\mu\nu,c} =
\varepsilon_{\mu\nu cd} \eta^d
$$
By analogy with the $d=3$ case for gravity one can easily construct
an example of (rather exotic) interaction by adding to the Lagrangian
and to the gauge transformation laws additional terms:
\begin{equation}
{\cal L}_1 = \frac{\kappa}{4} \varepsilon^{\mu\nu\alpha\beta}
\varepsilon_{abcd} f_\mu{}^a f_\nu{}^b f_\alpha{}^c \delta_\beta{}^d
\qquad \delta_1 \Phi_{\mu\nu,c} = \frac{\kappa}{2} f_{[\alpha}{}^a
\delta_{\beta]}{}^b \varepsilon_{abcd} \eta^d
\end{equation}
Computing the commutator of two $\eta$-transformations we get:
$$
[\delta_1, \delta_2] \Phi_{\alpha\beta,c} = \partial_{[\alpha}
z_{\beta]c}, \qquad z_{\beta c} = \frac{\kappa}{2}
\varepsilon_{a\beta cd} \eta_1{}^a \eta_2{}^d
$$
so that the whole algebra of $z$ and $\eta$-transformations
closes.

We can proceed and consider next order interaction by introducing
quartic terms to the Lagrangian and quadratic ones to the gauge
transformation laws:
\begin{equation}
\Delta {\cal L} = - \frac{\kappa_2}{8} \varepsilon^{\mu\nu\alpha\beta}
\varepsilon_{abcd} f_\mu{}^a f_\nu{}^b f_\alpha{}^c f_\beta{}^d \qquad
\delta \Phi_{\alpha\beta,c} = \kappa_2 f_\alpha{}^a f_\beta{}^b
\varepsilon_{abcd} \eta^d
\end{equation}
In this, we obtain additional terms for the commutator of two
$\eta$-transformations:
\begin{equation}
[\delta_1, \delta_2] \Phi_{\alpha\beta,c} = \kappa_2
\partial_{[\alpha} \tilde{z}_{\beta]c} - \kappa_2 \partial_{[\alpha}
f_{\beta]}{}^a \varepsilon_{abcd} \eta_1{}^a \eta_2{}^d
\end{equation}
where $\tilde{z}_{\beta c} = f_\beta{}^b \varepsilon_{abcd}
\eta_1{}^a \eta_2{}^d$.

By adjusting the values of two arbitrary coupling constants $\kappa$
and $\kappa_2$ and adding necessary linear terms to the Lagrangian the
whole model could be written in the formally covariant form:
\begin{equation}
{\cal L} = \varepsilon^{\mu\nu\alpha\beta} [ \frac{1}{32}
\varepsilon_{abcd} E_\mu{}^a E_\nu{}^b E_\alpha{}^c E_\beta{}^d -
E_\mu{}^a T_{\nu\alpha\beta}{}^a ]
\end{equation}
where $E_\mu{}^a = \delta_\mu{}^a + \kappa f_\mu{}^a$. This Lagrangian
is invariant under the following local gauge transformations:
\begin{equation}
\delta E_\mu{}^a = \partial_\mu \eta^a, \qquad \delta
\Phi_{\alpha\beta,c} = \partial_{[\alpha} z_{\beta]c} + \frac{1}{2}
E_\alpha{}^a E_\beta{}^b \varepsilon_{abcd} \eta^d
\end{equation}
the commutator of two $\eta$-transformations having the form:
\begin{equation}
[\delta_1, \delta_2] \Phi_{\alpha\beta,c} = \frac{1}{2}
\partial_{[\alpha} \tilde{z}_{\beta]c} - \frac{1}{2}
\partial_{[\alpha} E_{\beta]}{}^b \varepsilon_{abcd} \eta_1{}^a
\eta_2{}^d
\end{equation}
and $\tilde{z}_{\beta c} = E_\beta{}^b \varepsilon_{abcd}
\eta_1{}^a \eta_2{}^d$.

\section{$T_{[\mu\nu\alpha],\beta}$ tensor}

Our next example --- mixed tensor $T_{\mu\nu\alpha,\beta}$
antisymmetric on first three indices. This case is very much similar
to the previous one so we will be brief. Usual description uses this
tensor with additional condition $T_{[\mu\nu\alpha,\beta]} = 0$ and
two gauge transformations with parameters $\chi_{[\mu\nu],\alpha}$
such that $\chi_{[\mu\nu,\alpha]} = 0$ and $\eta_{[\mu\nu\alpha]}$. In
this, to construct minimal gauge invariant expression one needs at
least two derivatives. Let us abandon additional condition
$T_{[\mu\nu\alpha,\beta]} = 0$ and simultaneously combine to gauge
transformation into unconstrained one:
\begin{equation}
\delta T_{\mu\nu\alpha,\beta} = \partial_{[\mu}
\chi_{\nu\alpha],\beta}
\end{equation}
Then it is trivial task to construct gauge invariant object of the
first order:
$$
E_{\mu\nu\alpha\beta,\rho} = \partial_{[\mu} T_{\nu\alpha\beta],\rho}
$$
Now if we consider the most general (gauge invariant) Lagrangian
quadratic in $E$-tensor and require that this Lagrangian be invariant
under the local shifts
$\delta T_{\mu\nu\alpha,\beta} = \eta_{\mu\nu\alpha\beta}$
we obtain:
\begin{equation}
{\cal L} = \frac{1}{24} E^{\mu\nu\alpha\beta,\rho}
E_{\mu\nu\alpha\beta,\rho} + \frac{1}{18} E^{\mu\nu\alpha\beta,\rho}
E_{\mu\nu\alpha\rho,\beta} - \frac{2}{9} E^{\mu\nu\alpha}
E_{\mu\nu\alpha}
\end{equation}

Let us make a transition to the first order formalism. First of all we
introduce auxiliary field $\Omega_{\mu,[\nu\alpha\beta\rho]}$
antisymmetric on the last four indices which will play the role of
gauge field for the $\eta$-transformations. Then we construct a first
order Lagrangian
\begin{equation}
{\cal L}_I = - \frac{2}{9} \Omega^{\rho,\mu\nu\alpha\beta}
\Omega_{\mu,\rho\nu\alpha\beta} + \frac{2}{9} \Omega^{\mu\nu\alpha}
\Omega_{\mu\nu\alpha} - \frac{1}{9} \Omega^{\rho,\mu\nu\alpha\beta}
E_{\mu\nu\alpha\beta,\rho} - \frac{4}{9} \Omega^{\mu\nu\alpha}
E_{\mu\nu\alpha}
\end{equation}
which is invariant not only under the $\chi$-transformations (), but
also under
$$
\delta \Omega_\rho{}^{\mu\nu\alpha\beta} = \partial_\rho
\eta^{\mu\nu\alpha\beta}, \qquad \delta T_{\mu\nu\alpha,\beta} =
\eta_{\mu\nu\alpha\beta}
$$

If one solves the algebraic equation of motion for the $\Omega$ field
one get
\begin{equation}
\Omega_{\rho,\mu\nu\alpha\beta} = - \frac{3}{4}
E_{\mu\nu\alpha\beta,\rho} - \frac{1}{4} E_{\rho[\nu\alpha\beta,\mu]}
\end{equation}
In this, substitution of this expression back into the first order
Lagrangian () results exactly in the second order Lagrangian ().
As in the previous case by dividing all indices into the "local" and
"world" ones the first order Lagrangian () could be rewritten in the
simple and convenient form:
\begin{equation}
{\cal L}_I = \frac{2}{9} \left[ \left\{ \phantom{|}^{\mu\nu}_{ab}
\right\} \Omega_\mu{}^{acde} \Omega_\nu{}^{bcde} - \frac{1}{48}
\left\{ \phantom{|}^{\mu\nu\alpha\beta\gamma}_{abcde} \right\}
\Omega_\mu{}^{abcd} E_{\nu\alpha\beta\gamma}{}^e \right]
\end{equation}

This time our construction works for the space-time dimensions
$d \ge 5$ only. In the minimal $d=5$ case one can introduce dual
variables:
\begin{equation}
f_\mu{}^a = \frac{1}{24} \varepsilon^{abcde} \Omega_\mu{}^{bcde},
\qquad \eta^a = \frac{1}{24} \varepsilon^{abcde} \eta_{bcde}
\end{equation}
Then the Lagrangian takes the form:
\begin{equation}
{\cal L} = \frac{4}{3} \left\{ \phantom{|}^{\mu\nu}_{ab} \right\}
f_\mu{}^a f_\nu{}^b - \frac{1}{9}
\varepsilon^{\mu\nu\alpha\beta\gamma} f_\mu{}^a
E_{\nu\alpha\beta\gamma}{}^a
\end{equation}
while the gauge transformations leaving it invariant look like:
$$
\delta f_\mu{}^a = \partial_\mu \eta^a, \qquad \delta
T_{\mu\nu\alpha,a} = \varepsilon_{\mu\nu\alpha ab} \eta^b
$$
It is not hard to give an example of interaction for this case by
adding to the Lagrangian and gauge transformation laws new terms:
\begin{equation}
{\cal L}_1 =\kappa \left\{ \phantom{|}^{\mu\nu\alpha}_{abc} \right\}
f_\mu{}^a f_\nu{}^b f_\alpha{}^c, \qquad \delta_1
T_{\alpha\beta\gamma,a} = \kappa
\varepsilon_{abc[\alpha\beta} f_{\gamma]}{}^b \eta^c
\end{equation}
Again this Lagrangian looks very simple in the first order formalism,
but it corresponds to highly non-linear second order Lagrangian with
non limited number of derivatives.

\section{$R_{[\mu\nu],[\alpha\beta]}$ tensor}

In this section we consider more interesting and less evident example
--- tensor $R_{[\mu\nu],[\alpha\beta]}$. Usual description is based on
the tensor with constraint $R_{\mu[\nu,\alpha\beta]} = 0$ and gauge
transformations with a parameter $\chi_{\mu,[\alpha\beta]}$ also
constrained by $\chi_{[\mu,\alpha\beta]} = 0$. We abandon both
constraints and consider arbitrary $R_{[\mu\nu],[\alpha\beta]}$ tensor
with gauge transformations
$\delta R_{\mu\nu,\alpha\beta} = \partial_\mu \chi_{\nu,\alpha\beta}
- \partial_\nu \chi_{\mu,\alpha\beta} $, where
$\chi_{\mu,\alpha\beta}$ antisymmetric on the last two indices. Then
it is trivial to get gauge invariant "field strength":
$$
T_{[\mu\nu\alpha],[\beta\gamma]} = \partial_{[\mu}
R_{\nu\alpha],\beta\gamma}
$$
Now we consider the most general Lagrangian quadratic in
$T_{\mu\nu\alpha,\beta\gamma}$ and require that it has to be invariant
under the local shifts:
$$
R_{\mu\nu,\alpha\beta} \rightarrow R_{\mu\nu,\alpha\beta} +
\eta_{\mu,\nu\alpha\beta} - \eta_{\nu,\mu\alpha\beta}
$$
where $\eta_{\mu,\nu\alpha\beta}$ antisymmetric on the last three
indices. We obtain:
\begin{eqnarray}
{\cal L} &=& \frac{1}{6} T^{\mu\nu\alpha,\beta\gamma}
T_{\mu\nu\alpha,\beta\gamma} + \frac{1}{2}
T^{\mu\nu\alpha,\beta\gamma} T_{\mu\nu\beta,\alpha\gamma} +
\frac{1}{2} T^{\mu\nu\alpha,\beta\gamma} T_{\mu\beta\gamma,\nu\alpha}
- \nonumber \\
 && - \frac{3}{2} T^{\mu\nu,\alpha} T_{\mu\nu,\alpha} - 3
 T^{\mu\nu,\alpha} T_{\mu\alpha,\nu} + \frac{3}{2} T^\mu T_\mu
\end{eqnarray}

To make a transition to first order formalism we introduce auxiliary
field $\Omega_{[\mu\nu],[\alpha\beta\gamma]}$ antisymmetric on the
first two as well as last three indices which will play a role of
gauge
field for the $\eta$-transformations. The following first order
Lagrangian:
\begin{eqnarray}
{\cal L} &=& - \frac{3}{2} \Omega^{\mu\nu,\alpha\beta\gamma}
\Omega_{\alpha\beta,\mu\nu\gamma} + 6 \Omega^{\mu,\nu\alpha}
\Omega_{\nu,\mu\alpha} - \frac{3}{2} \Omega^\mu \Omega_\mu + \nonumber
\\
 && + \Omega^{\mu\nu,\alpha\beta\gamma} T_{\alpha\beta\gamma,\mu\nu}
 + 6 \Omega^{\mu,\nu\alpha} T_{\nu\alpha,\mu} + 3 \Omega^\mu T_\mu
\end{eqnarray}
is invariant not only (by construction) under the local $\chi$-
transformations, but under the local transformations
$$
\delta \Omega_{\mu\nu,\alpha\beta\gamma} = \partial_\mu
\eta_{\nu,\alpha\beta\gamma} - \partial_\nu
\eta_{\mu,\alpha\beta\gamma} \qquad
\delta R_{\mu\nu,\alpha\beta} =
\eta_{\mu,\nu\alpha\beta} - \eta_{\nu,\mu\alpha\beta}
$$
as well. By solving algebraic equation of motion for the $\Omega$
field one obtains:
\begin{equation}
\Omega_{\mu\nu,\alpha\beta\gamma} = \frac{1}{3}
T_{\alpha\beta\gamma,\mu\nu} + \frac{1}{6} [ T_{\mu\beta\gamma,\alpha
\nu} - T_{\nu\beta\gamma,\alpha\mu} + (\alpha,\beta,\gamma)] +
\frac{1}{3} [ T_{\mu\nu\alpha,\beta\gamma} + (\alpha,\beta,\gamma) ]
\end{equation}
Substituting this expression back into the first order Lagrangian one
reproduces exactly the second order Lagrangian.

As in all previously considered cases if we divide all indices into
the "local" and "world" ones it is possible to rewrite our Lagrangian
in the following form:
\begin{equation}
{\cal L} = - \frac{3}{8} \left\{
\phantom{|}^{\mu\nu\alpha\beta}_{abcd} \right\}
\Omega_{\mu\nu}{}^{abe} \Omega_{\alpha\beta}{}^{cde} + \frac{1}{12}
\left\{ \phantom{|}^{\mu\nu\alpha\beta\gamma}_{abcde} \right\}
\Omega_{\mu\nu}{}^{abc} T_{\alpha\beta\gamma}{}^{de}
\end{equation}
This time our construction also works for the space-time dimensions
$d \ge 5$ only.

\section*{Conclusion}

Thus we have managed to construct first order formalism for the three
examples of mixed symmetry high spin fields. In all three cases the
formulation is very much similar to the usual tetrad formalism in
gravity so that the Lagrangians have very simple and suggestive form
and gauge invariance is almost evident. We hope that such formalism
could be useful in investigations of possible interactions for such
fields.


\newpage

\begin{thebibliography}{99}
\bibitem{Cur86}
T.~Curtright, {\it "Generalized gauge fields",} Phys. Lett. {\bf B165}
(1986) 304.
\bibitem{CF86}
T.~Curtright, P.~G.~O.~Freund, {\it "Massive dual fields",} Nucl.
Phys. {\bf B172} (1986) 413.
\bibitem{AKO86}
C.~S.~Aulakh, I.~G.~Koh, S.~Ouvry, {\it "Higher spin fields with mixed
symmetry",} Phys. Lett. {\bf B173} (1986) 284.
\bibitem{LM86}
J.~M.~Labastida, T.~R.~Morris, {\it "Massless mixed symmetry bosonic
free fields",} Phys. Lett. {\bf B180} (1986) 101.
\bibitem{Lab89}
J.~M.~Labastida, {\it "Massless particles in arbitrary representations
of the Lorentz group",} Nucl. Phys. {\bf B322} (1989) 185.
\bibitem{GK98}
J.~A.~Garcia, B.~Knaepen, {\it "Couplings between generalized gauge
fields",} Phys. Lett. {\bf B441} (1998) 198, hep-th/9807016.
\bibitem{BPT01}
C.~Burdik, A.~Pashnev, M.~Tsulaia, {\it "On the mixed symmetry
irreducible representations of the Poincare group in the BRST
approach",} Mod. Phys. Lett. {\bf A16} (2001) 731, hep-th/0101201.
\bibitem{Zin02}
Yu.~M.~Zinoviev {\it "On Massive Mixed Symmetry Tensor Fields in
Minkowski space and (A)dS",} hep-th/0211233.
\bibitem{BB02}
X.~Bekaert, N.~Boulanger, {\it "Tensor gauge fields in arbitrary
representations of GL(D,R) : duality and Poincare lemma", }
hep-th/0208058.
\bibitem{MH02}
P.~de~Medeiros, C.~Hull, {\it "Exotic tensor gauge theory and
Duality", } Com. Math. Phys. {\bf 235} (2003) 255, hep-th/0208155.
\bibitem{BG00}
N.~Boulanger, L.~Gualteri, {\it "An exotic theory of massless spin-two
fields in three dimensions",} Class. Quant. Grav. {\bf 18} 1485,
hep-th/0012003.
\bibitem{BCNS02}
C.~Bizdadea, E.~M.~Cioroianu, I.~Negru, S.~O.~Saliu, {\it "Lagrangian
interactions within a special class of covariant mixed-symmetry type
tensor gauge fields",} hep-th/0211158.
\bibitem{BCCSS03}
C.~Bizdadea, C.~C.~Ciobirca, E.~M.~Cioroianu, S.~O.~Saliu, S.~C.~
Sararu, {\it "Interacting mixed-symmetry type tensor gauge fields of
degrees two and three: a four-dimensional cohomological approach",}
hep-th/0303079.
\end{thebibliography}
\end{document}